\documentclass[aps,prl,twocolumn,superscriptaddress,floatfix]{revtex4-1}
\usepackage{graphicx,amsmath,subfigure,epsfig,psfrag,verbatim}
\begin{document}
\title{Decoding Spatial Complexity in Strongly Correlated Electronic Systems}

\author{E.~W.~Carlson}
\affiliation{Department of Physics, Purdue
University, West Lafayette, IN 47907, USA}
\author{S.~Liu}
\affiliation{Department of Physics, Purdue University, West
Lafayette, IN 47907, USA}
\author{B.~Phillabaum}
\affiliation{Department of Physics, Purdue University, West
Lafayette, IN 47907, USA}
\author{K.~A.~Dahmen}
\affiliation{Department of Physics, University of Illinois, Urbana-Champaign, IL, 50000, USA}
\date{\today}

\begin{abstract}
Inside the metals, semiconductors, and magnets of our everyday experience, electrons are uniformly distributed throughout the material.  By contrast, electrons often form clumpy patterns inside of strongly correlated electronic systems (SCES) such as colossal magnetoresistance materials and high temperature superconductors.  In copper-oxide based high temperature superconductors, scanning tunneling microscopy (STM) has detected an electron nematic on the surface of the material, in which the electrons form nanoscale structures which break the rotational symmetry of the host crystal.   These structures may hold the key to unlocking the mystery of high temperature superconductivity in these materials, but only if the nematic also exists throughout the entire bulk of the material.
Using newly developed methods  for decoding these surface structures,
 we find that the nematic indeed persists throughout the bulk of the material.  We furthermore find that the intricate pattern formation is set  by a delicate balance among disorder, interactions, and material anisotropy, leading to a fractal nature of the cluster pattern.  The methods we have developed can be extended to many other surface probes and materials, enabling surface probes to determine whether surface structures are confined only to the surface, or whether they extend throughout the material.
\end{abstract}

\maketitle

There is growing experimental evidence that many strongly correlated electronic systems such as
nickelates, cuprates, and manganites exhibit some degree of local inhomogeneity,\cite{dagotto-science,Lai:2010p865,
%Wise:2009p866,
Qazilbash:2007p615,Sun:2007p867,
%dagotto-cmr,Mn-phssep-1,Mn-phssep-2,Mn-phssep-3,
tranqnature,
%hoffman,
%checknccoc,uemura2000,hammelwipeout,gaplessbscco,seamus-glass,
%cnl,dhleeloops,leeloops,testso5,
zaanengunnarsson,fratini-powerlaw}
{\em i.e.}, nanoscale variations in the local electronic properties.
Describing the electronic behavior of these materials involves several degrees of freedom, including orbital, spin, charge, and lattice degrees of freedom.
Disorder only compounds the problem. Not only can
disorder destroy
phase transitions, leaving mere crossovers in the wake, it can fundamentally alter ground
states, sometimes forbidding long range order. Especially in systems where different physical
tendencies compete,
disorder can act as nucleation points for competing ground states.

One approach to disentangling disorder from the fundamental
correlations induced by strong electron interactions is to put
resources toward developing cleaner samples.    While this approach
is laudable and has led to many key insights and advances in
strongly correlated electronic systems, it is also labor intensive
and expensive.      Even the cleanest sample, when stored over time at
finite temperature, will acquire a thermodynamically required
concentration of defects.  Moreover, in some sense, disorder is
intrinsic to the correlated phases, since in most systems the
phases of interest happen upon chemical doping, which
necessarily introduces disorder.  Especially in cuprates,
this drive toward cleaner samples or even toward controlling disorder
in order to understand the intrinsic electronic states
may not be necessary, since even ``dirty'' samples that have
not undergone strict preparation protocols still exhibit the salient feature
of superconductivity.\cite{ornl-highschool}
Indeed, in any high temperature superconductor,
because the pairing scale must also be high, an understanding
of the short-distance physics ({\em i.e.} within a few
coherence lengths of the superconductivity) should be
sufficient to understand the origin of pairing.\cite{concepts}
In this sense, long range order of a proposed pseudogap phase
is neither necessary to  produce superconductivity
nor is it necessary in order to understand the superconductivity.

Ultimately, the interplay between many degrees of freedom,  strong correlations,
and disorder can lead to  a hierarchy of
length scales over which the resulting physics must be described.\cite{dagotto-science}
While such electronic systems are highly susceptible to pattern formation at the nanoscale,
unfortunately most of our theoretical and experimental tools are designed for understanding
and detecting homogeneous phases of matter.
Therefore, there is a critical need to design and develop new ways of understanding, detecting, and characterizing
electronic pattern formation in strongly correlated electronic systems at the nanoscale,
especially in the presence of severe disorder effects.
Such theoretical guidance will enable more direct contact between theory and experiment
in a number of materials, and provide a path forward for
understanding ``disputed'' regions of phase diagrams of strongly correlated materials.

\begin{figure}
\centering
\includegraphics[width=0.95\columnwidth]{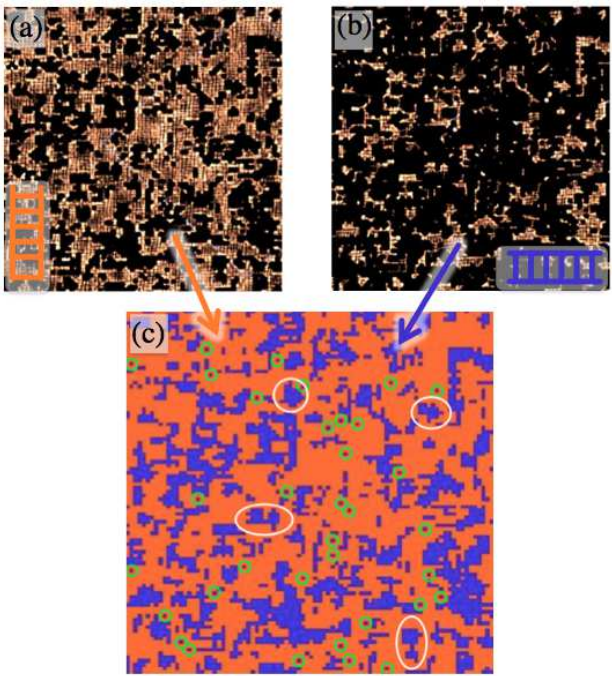}%
\caption{
Mapping of STM data to Ising nematic variables.
(a) Masked image\cite{phillabaum,kohsaka-science-2007}
of R-map of Dy-Bi2212
showing the regions of the R-map with vertically
aligned nematic domains.  (b) The complement
to panel (a), showing horizontally aligned nematic domains.
(c) Mapping to the corresponding Ising geometric clusters,
showing several small clusters (circled in green);
a smaller number of medium-sized clusters
(representative clusters circled in white);
and a single (orange) cluster which spans the entire field of view.
}
\label{fig:clusters}
\end{figure}

In this paper, we focus on detecting electron nematics and
other  electronic phases which break the
rotational symmetry of the host crystal.  Such phases have
been proposed and/or observed in a variety of materials
and contexts, including
Sr$_3$Ru$_2$O$_7$\cite{Borzi:2007p642},
GaAs/Al$_x$Ga$_{1- x}$As heterostructures in field\cite{Cooper:2002p778,Du:1999p789},
and a subset of cuprate superconductors\cite{Fradkin:2010p446,Ando:2002p780,Hinkov:2004p130,Daou,Howald:2003p786,kohsaka-science-2007,Lawler:2010p639}
 such as YBa$_2$Cu$_3$O$_{6+x}$\cite{Ando:2002p780,Hinkov:2004p130,Daou},
and Bi$_2$Sr$_2$CaCu$_2$O$_{8+x}$\cite{Howald:2003p786,kohsaka-science-2007,Lawler:2010p639},
as well as the iron arsenic based superconductor Ca(Fe$_{1-x}$Co$_x$)$_2$As$_2$\cite{Chuang:2010p643}.
The state has been proposed to exist in many more systems, such as
AlAs heterostructures, the Si(111) surface, elemental bismuth,
and both single layer and bilayer graphene.\cite{Abanin:2010p646,Vafek:2010p647,Rasolt:1986p794,Fradkin:2010p446}

We have proposed three approaches which, rather than shying
away from disorder, use disorder to advantage in order to
detect and characterize mesoscale and multi scale order
in electronic systems (such as electron nematics) which break the rotational symmetry of the host crystal:
(1)  Extracting critical exponents from observed multi scale pattern formation in image data via
cluster analyses.\cite{fratini-powerlaw,phillabaum};
(2)  Manifestations of nonequilibrium behavior such as hysteresis\cite{detecting-nematics,degiorgi-2014}; and
(3) Noise characteristics\cite{Bonetti:2004dr,rfim-prl}.

Method \# 1 (Cluster Analyses) above is the subject of this paper,
and the key insight is that near criticality, most
physical quantities display power law behavior on length
scales smaller than an appropriately defined correlation length.
In order to make the connection with image data, this requires
translating the geometric patterns into critical exponents,
as described below.\cite{fratini-powerlaw,phillabaum}
Method \#2 (Hysteresis) relies on the extreme critical slowing down
accompanying certain classes of quenched disorder.\cite{fisher-exponential-barriers}
For systems in which the quenched disorder is of the
{\em random field} type (see Eqn.~\ref{eqn:model} below),
hysteresis is a prominent and robust feature,
which means that hysteresis can be a good diagnostic tool
for order parameters which couple to material disorder
via a random field mechanism.\cite{rfim-prl}
In this case, we have proposed using hysteresis in order
to detect disordered electron nematics, even ones which
never fully order but exhibit only local nematic order.
The key insight is to field cool in an orienting field
(such as uniaxial pressure), and measure any macroscopic
response function which is sensitive to nematic order
(such as anisotropic resistivity).  Through specific
field cooling and orientational field switching protocols
as described in Ref.~\cite{detecting-nematics},
the presence of a disordered electron nematic can be
revealed experimentally.\cite{degiorgi-2014}
Method \#3 (Noise Characteristics) above
concerns another manifestation of the slow time dynamics associated with
random field disorder.
For example, very slow telegraph noise was observed in
the transport properties of a
YBCO nanowire in the pseudogap phase\cite{Bonetti:2004dr},
consistent with the expected resistivity fluctuations
of mesoscale electron nematic patches
thermally switching their orientation.\cite{rfim-prl}.

Patterns of unidirectional domains have been detected
on the surface of cuprate superconductors.\cite{kohsaka-science-2007}
In Fig.~\ref{fig:clusters}(a) and (b), we show the patterns of vertically oriented
and horizontally oriented stripe domains, respectively,
derived from a local Fourier transform of the R-map
of STM on Dy-Bi2212, from Supplementary Fig. S3
of Ref.~\cite{kohsaka-science-2007}, as detailed in Ref.~\cite{phillabaum}.
Fig.~\ref{fig:clusters}(c) shows the corresponding Ising cluster map,
where orange represents vertically aligned clusters,
and blue represents horizontally aligned clusters.
From the figure, it is evident that there is one large
system-spanning (orange) cluster.
There are also several medium-sized clusters
(some are circled in white in Fig.~\ref{fig:clusters}(c)),
as well as many small clusters (circled in green
in Fig.~\ref{fig:clusters}(c)).

The clusters display structure on all length
scales within the field of view.
In addition, as we will show below, the boundaries
of the clusters are fractal, and the sizes
of the clusters are power law distributed.
These are all features which point to
the pattern formation being driven by proximity to a
critical point.
If there is an underlying critical point driving
the pattern formation, then critical exponents
are encoded in the image.\cite{fratini-powerlaw,phillabaum}.
For example, the number of clusters $D$ of a particular
size $S$ is power-law distributed in this image,
$D(S) \propto S^{-\tau}$, with a power set by
the Fisher exponent $\tau$.
In addition, the fractal geometric structure
of the clusters can be quantified as
the hull fractal dimension, $d_h$,
and the volume (interior) fractal dimension, $d_v$
of clusters.
By studying the orientational analogue
of the spin-spin correlation function,
the anomalous dimension $\eta_{||}$ can also
be extracted from the image.

%*** ADDRESS THESE POINTS ***
%- Violation of hyperscaling does not necessarily mean quantum
%- Violation of hyperscaling in RFIM does not require fine-tuning.
%- Does this also point to optimal inhomogeneity, fractal inhomogeneity,
%favoring HTSC?\cite{bianconi}

Relating
these critical exponents
to a particular fixed point
requires a model.
Near a critical point, the correlation length grows to become the
dominant length scale, and it is possible to map the real physical
system to a coarse-grained model with the same universal features.
Starting from the cluster map in Fig.~\ref{fig:clusters}(c), we assign
Ising variable $\sigma = 1$ to vertical domains, and $\sigma = -1$  to
horizontal domains.\cite{phillabaum}
We furthermore incorporate disorder into the
model in the following way:  In any given region of the sample,
dopant disorder locally breaks the rotational symmetry of the
Cu-O plane.  (The same is true of other sources of quenched disorder.)
This locally favors one orientation of the nematic over the other.
Thus, quenched disorder acts as a random field on the
local director of the Ising nematic.\cite{rfim-prl}
The model may be stated as
\begin{eqnarray}
H&=&-\sum_{\langle ij\rangle_\parallel}J_{||} \sigma_i \sigma_j -\sum_{\langle
ij\rangle_{\perp}}J_{\perp} \sigma_i
\sigma_j \nonumber \\
&-&\sum_i h_i\sigma_i~, \label{eqn:model}
\end{eqnarray}
%EC 8/9/13 Define the sum
where the sum runs over the coarse-grained regions (Ising sites)
consisting of a cubic lattice, chosen with spacing
comparable to the resolution of the image(s) to be studied. The tendency for
neighboring regions to be of like character is modeled as a nearest
neighbor ferromagnetic interaction $J>0$.
The layered structure of the material is captured by
the in-plane coupling $J_{||}$ being  larger
than the coupling between planes $J_{\perp}$.
Ultimately, the criticality of such a quasi-two-dimensional system
is controlled by a three dimensional fixed point
for any finite $J_{\perp}$.
However, in a strongly layered system such as the cuprates
where $J_{\perp} << J_{||}$,
it is possible to observe a drift from two dimensional
to three dimensional exponents when
observing a finite field of view.\cite{Zachar:2003eq}

%At the order parameter level, there are two broad classes of
%disorder: local energy density disorder (which we incorporate as
%random bond disorder), and random field disorder \cite{cardy-book}.
%Random bond disorder is included through the term $\delta J_{ij}$,
%and $h_i$ represents random field disorder, which is chosen from a
%gaussian probability distribution centered about zero, with variance
%$\Delta$. The field $h$ represents a generalized external field
%which couples with the local Ising variables.
%

There are six critical fixed points which can arise from
the model of Eqn.~\ref{eqn:model}:
In the limit of zero disorder strength,  the phase
transition from disordered to long-range ordered nematic
is controlled by the two-dimensional clean Ising model (C-2D)
if $J_{\perp} = 0$, or by the three-dimensional
clean Ising model (C-3D) for any nonzero coupling between planes.
Random field disorder is relevant, and so the presence of
any finite amount of random field disorder $\Delta$
shifts the universality class to either the two-dimensional
random field Ising model (RF-2D) or the
three-dimensional random field Ising model (RF-3D).\cite{phillabaum}
Note that quenched material disorder can also give rise to
randomness in the coupling strengths $J_{\perp}$ and
$J_{||}$, also known as random bond disorder.
However, in the presence of both random bond and random field disorder,
the critical behavior is always controlled by the random field fixed point.
For completeness, we also consider the possibility
that the observed local orientations are not arising from
an interacting model, which corresponds to the percolation
fixed points which occur at the infinite temperature
limit of Eqn.~\ref{eqn:model} as a function of applied
orienting field.  These are the two-dimensional
and three-dimensional uncorrelated percolation points,
P-2D and P-3D, respectively.

\begin{figure}
  \centering
  \includegraphics[width=0.95\columnwidth]{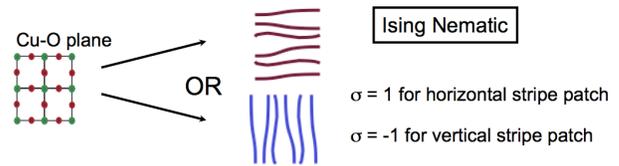}
  \caption{
An electron nematic breaks the rotational symmetry
of the host crystal, in this case from C4 to C2 symmetry.
The electron nematic then aligns either ``vertically'' or
``horizontally.''  We assign Ising variables $\sigma = -1$
and $\sigma = +1$, respectively.
  }
  \label{pd-3D}
\end{figure}

\begin{figure}
  \centering
  \includegraphics[width=0.95\columnwidth]{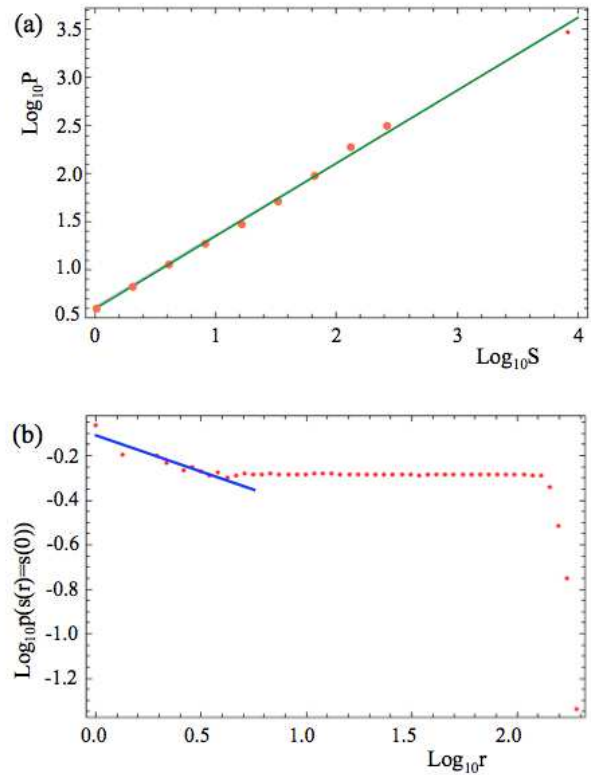}%
  \caption{
  Scale-free behavior of electron nematic clusters in Dy-Bi2212.
  (a) For each cluster size $S$ (defined as the number of sites $S$ in the cluster),
  the average perimeter is plotted.  Scaling is evident throughout
  the entire field of view.
 Green line is a linear fit, yielding the ratio of fractal
 dimensions as described in the text.
  (b) The probability $p$ that two spins a distance $r$ apart are aligned.
The blue line is a linear fit, as described in the text.
  In both panels, logarithmic binning has been used,
  which is a standard technique for power law analysis.\cite{Newman:2005vk}
  }
\label{fig:powerlaws}
\end{figure}

As shown in Fig.~\ref{fig:clusters}(c),
there is one large spanning cluster,
and there are several
medium-sized clusters, and even more
small clusters.  By counting the number
of clusters $D$ of each size $S$
(where $S$ is the number of Ising sites
in each cluster), one can construct
the cluster size distribution $D(S)$.
Near a critical point, this quantity exhibits
power law scaling, as $D(S) \propto S^{-\tau}$.
However, it is known that near criticality,
the scaling function which forms the prefactor
for the power law has a pronounced bump\cite{perkovic},
at least for the 3D random field fixed point.
Therefore,  a finite-size field of view
is expected to underestimate the true
value of $\tau$.  In future experiments, larger fields
of view can mitigate this effect. Within the
field of view available, we find that a straightforward
power law fit yields $\tau = 1.71 \pm 0.07$.

From the cluster structure in Fig.~\ref{fig:clusters},
one can see that the clusters themselves are not compact.
Rather, they are ethereal, even gossamery\cite{laughlin}
in nature.
Indeed, both the boundaries and the interiors of the
clusters are fractal in nature.
In Fig.~\ref{fig:powerlaws}(a),
we show a log-log plot of the average perimeter $P$
of clusters of each size $S$.  As with the cluster size distribution,
a robust power law emerges throughout the entire field of view.
By comparing the perimeter and cluster sizes, the ratio
of fractal dimensions
$P \propto S^{d*_h/d*_v}$ can be extracted,
where $d_h$ and $d_v$ denote the hull and volume
fractal dimensions, respectively.
(Here, the asterisk denotes the fact
that only a 2D slice of the clusters is experimentally accessible,
and therefore a corresponding
geometric factor must be applied before comparing
directly with 3D models.\cite{phillabaum})
The ratio of fractal dimensions thus obtained is
$d*_h/d*_v = 0.78 \pm 0.01.$

In Fig.~\ref{fig:powerlaws}(b), we plot the probability
$p(r)$ that two {pseudospins} $\sigma = \pm 1$ a distance $r$ apart are aligned.
This is linearly related to the spin-spin correlation function
$g(r) \propto p(r)$.
(In the physical system, the spin-spin correlation function
corresponds to the orientation-orientation correlation function
of the director of the electron nematic.)
The spin-spin correlation function becomes
power law near a critical point, $g(r) \propto 1/r^{d-2+\eta_{||}}$.
Here, we denote by $\eta_{||}$ the anomalous dimension $\eta$
at the surface of a material.
As can be seen
in the figure, this function is at most weakly power law
in the data, with less than a decade of scaling.
As such, this is the least reliable critical exponent
extracted from the cluster analysis, yielding a value
$d-2+\eta_{||} = 0.8 \pm 0.3$.

\begin{centering}
\begin{widetext}
\begin{table}[htbp]
%  \centering
  \begin{tabular}{@{} |c|c|c|c|c|c|c||c| @{}}
    \hline
    {\rm Exponent~$\downarrow$; Model~$\rightarrow$}
 & {\rm C-2D} &{\rm C-3D} & {\rm P-2D} &{\rm P-3D} & {\rm RF-2D} & {\rm RF-3D } & {\rm Dy-Bi2212}\\
    \hline
    $\tau$ & 2.076 &2.21 & 2.02 & 2.18 & 2.0 & $2.02 \pm 0.03$ & $1.71 \pm 0.07$ \\
    $d*_h/d*_v$ & 0.71 & - & 0.92 & 0.74 & .92 & 0.57 & $0.78 \pm 0.01$ \\
    $d-2+\eta_{||}$ & 0.25 & 2.54 & 0.207 & 0.93 & 1 & 0.336 & $ 0.8 \pm 0.3$ \\
    \hline
  \end{tabular}
  \caption{
Comparing critical exponents derived from R-Map STM data on
Dy-Bi2212 to theoretical values of critical fixed points of the
model in Eqn.~\ref{eqn:model}.
Critical fixed points considered include
the clean two-dimensional and three-dimensional Ising models
(C-2D and C-3D, respectively); uncorrelated percolation
in two dimensions (P-2D); and the two-dimensional
and three-dimensional random field Ising models
(RF-2D and RF-3D, respectively).
Theoretical values of fixed points are
from Ref.~\cite{phillabaum} and references therein.
}
  \label{tab:label}
\end{table}
\end{widetext}
\end{centering}

Table~\ref{tab:label} shows a comparison between the
critical exponents derived from the observed
unidirectional electronic clusters in Dy-Bi2212 and
theoretical values from critical fixed points of Eqn.~\ref{eqn:model}.
Note that for layered clean and random field Ising models,
it is possible to observe a drift from 2D to 3D exponents
in going from smaller to larger fields of view.\cite{Zachar:2003eq}
However, no such dimensional crossover makes sense
when considering uncorrelated percolation.

We now compare the data-derived exponents against theoretical models.
Note that the value of $\tau$ from the data is lower
than the theoretical value of every fixed point.
This is expected for a finite field of view in random field models,
where it is known that the cluster size distribution $D(S)$
has a pronounced scaling bump.\cite{perkovic}
Note also that there is not much variation in
the theoretical values of $\tau$
among the fixed points, so that while the presence
of a robust power law in $D(S)$ in the data is significant,
it is difficult in principle to determine {\em which}
fixed point could be responsible for the scale-free behavior
via this exponent.

By contrast, the anomalous exponent $d-2+\eta_{||}$ shows
a wide variation among fixed points, and can in principle
be a good value to distinguish among fixed points.
Unfortunately, the data-derived value has rather large
error bars due to a limited regime of scaling, and
so yields little information in this case.

Solid information can be gleaned by comparing the
effective ratio of fractal dimensions, $d*_h/d*_v$.
For 2D models, this corresponds to the bulk fractal dimensions,
$d*_h/d*_v = d_h/d_v$.  For 3D models, the bulk fractal
dimensions differ from those observed on a 2D slice
via geometrical factors, so that $d*_h/d*_v = 3 d_h/(4 d_v)$.\cite{phillabaum}
This ratio shows distinguishable variation among fixed points,
and the data-derived value has small error bars and exhibits
decades of scaling.  All of this means that this exponent
ratio is useful for distinguishing among the fixed points.
Note that the observed ratio is inconsistent with uncorrelated
2D percolation (P-2D), and we can rule out this fixed point as
the origin of the pattern formation.
Although the P-3D fixed point may appear to be a reasonable
match, other considerations rule this out as the origin of the cluster pattern.
First, this fixed point occurs when 31\% of domains point one direction,
and the rest point the other.\cite{stauffer-book}
Such an extreme value of net nematicity would surely have
been observed in macroscopic measurements on Dy-Bi2212,
which is not the case. Second, while P-3D corresponds to the
point at which geometric clusters percolate in a 3D system,
this is not the same as the point at which those clusters
percolate on a slice.  Rather, at the 3D percolation point
when {\em viewed on a 2D slice}, there is one large spanning
cluster with many small clusters, and no robust power law
behavior on the slice.  So, the P-3D point can also be ruled out
as the origin of the complex pattern formation.

This leaves the possibility of a dimensional crossover
from 2D to 3D behavior in either the clean
or random field Ising models.
The expectation in the literature is that there should
be no well-defined fractal dimension of geometric clusters
at C-3D, since in fact geometric clusters do not
exhibit power law behavior at the C-3D point.\cite{dotsenko-1995}
However, recent studies indicate that when viewed
{\em on a 2D slice}, geometric clusters do
exhibit power law behavior at C-3D.\cite{saberi-C3Dx}
The ratio of fractal dimensions on a 2D slice is not  known
in this case, and will be discussed in a future publication.\cite{future-C3Dx}

The possibility of a dimensional crossover from
2D to 3D exponents in a layered random field model
is consistent with all data-derived exponents,
and is the most likely source of the observed
scale-free pattern formation of the electron nematic.
In contrast with the clean model,
geometric clusters do exhibit fractal dimensions
and scale-free behavior at the RF-3D fixed point.
Furthermore, if this identification is correct, then the
clear prediction is that all data-derived exponents
should drift away from the RF-2D values
and closer to the RF-3D values upon increasing
the field of view.

Other considerations also point to random field behavior:
It has been previously shown that the slow
telegraph noise observed in transport on a YBCO
nanowire in the pseudogap regime\cite{Bonetti:2004dr}
is consistent with the mapping of
electron nematic domains in a host crystal to the
random field Ising model.\cite{rfim-prl}
This identification also serves to unify several
experiments, in that it offers a concrete
explanation for why certain materials display
long-range orientational stripe order, and others do not.
While true long-range electron nematic order is possible
in a real 3D system, it is completely forbidden in a 2D
system in the presence of any nonzero random
field disorder.  Thus, in a highly layered system such
as the cuprates, many samples are expected to
display no long-range order of the electron nematic,
although in a layered RFIM, nematic clusters can
grow quite large within the plane even if true  long-range order
is never achieved.\cite{Lawler:2010p639,Zachar:2003eq}

The RF-3D fixed point is a zero temperature
fixed point, which has implications for dynamics
as well as future experimental tests of the critical exponents.
First, it means that the entire finite-temperature phase transition boundary
in the layered model exhibits extreme critical slowing down.
With typical critical slowing down, the relaxation time of the system
diverges as a power law as criticality is approached,
$\tau_{\rm relax} \propto 1/|T-T_c|^{-\nu z}$.
However, the dynamics of the 3D random field Ising model
are even more extreme near criticality,
with the relaxation time diverging {\em exponentially}
as criticality is approached, $\tau_{\rm relax} \propto {\rm exp}[\xi^{\theta}]$
where $\xi$ is the spin-spin correlation length
(which here corresponds to the orientation-orientation
correlation length of the nematic director), and
$\theta$ is the violation of hyperscaling exponent,
which is nonzero at this fixed point.\cite{fisher-exponential-barriers}
Second, because $\theta \ne 0$ at a zero temperature
fixed point, hyperscaling relations of critical exponents (which involve
the dimension of the underlying phenomenon) must be modified.\cite{fisher-exponential-barriers}
Third, there is the question of whether fine-tuning is
required to see power law behavior associated with
the RF-3D fixed point.  In fact, partly because it is a zero
temperature fixed point with pronounced nonequilibrium effects,
there is a wide critical region associated with this fixed point.
For example, in the zero temperature 3D RFIM,
critical behavior with 2 decades of scaling can be observed
even 50\% away from the critical point.\cite{perkovic}

Finally, we comment on the implications of
multiscale behavior in cuprate superconductors.
It is not just the electron nematic which exhibits
fractal behavior in a bismuth-based cuprate,
but similar behavior has been noted in the
lanthanum family of cuprates as well.
The local density of oxygen interstitials in
LaSrCuO  follows a power law
at optimal doping.\cite{fratini-powerlaw}
In addition, theoretical studies have shown
that there is a Goldilocks type of optimal
inhomogeneity (neither too little nor too much)
which favors superconductivity
in a strongly correlated electronic system.\cite{kivelson-optimal-inhomogeneity,Loh:2007hw}
The presence of inhomogeneity on multiple length scales,
with robust power laws, in both the doping concentrations
and also directly in the electronic degrees of freedom
may point to the optimal inhomogeneity being fractal in nature.
Much like the construction of the Eiffel tower
incorporates elements of a scale-free iron latticework
in order to optimize structural stability given a certain
amount of iron to work with, high temperature superconductors
may benefit from scale-free organization of
electronic degrees of freedom in order to optimize
the superconducting transition temperature.\cite{bianconi-fractal-superconductivity}

In summary, we conclude that the complex, scale-free
pattern\cite{phillabaum} of nematic clusters observed at the surface
of Dy-Bi2212 via STM\cite{kohsaka-science-2007}
is controlled by the 3D random field Ising model fixed point.
That is, the ethereal cluster structure is due to a combination
of interactions between clusters and quenched disorder
due to material defects throughout the bulk of the material.
As such, the pattern formation is not merely a surface effect.
Rather, the nematic clusters form deep inside the material,
and intersect the surface.  While this analysis cannot distinguish
between true macroscopic long-range order of the electron nematic
and short-range order, we can conclude that there is
significant {\em multiscale} order in the system.
Indeed, because the pairing energy scale is high,
the pairing mechanism can arise from short-distance physics,
and the presence of large nematic clusters throughout the
bulk of the material is sufficient for superconducting pairing
to originate from the electron nematic.

\begin{acknowledgments}
We thank J.~Hoffman, S.~Kivelson, Y.~Loh, E.~Main, B.~Phillabaum, and C.-L.~Song for helpful
conversations. S.L. and E.W.C. acknowledge support from NSF
Grant No. DMR 11-06187.
K.A.D.
acknowledges support from NSF Grant No. DMR 10-05209 and NSF Grant
No. DMS 10-69224.
\end{acknowledgments}

\bibliography{bigbib,smectic,grants,rfim,spin,detectingnematics,rfim-mott,3-state-potts,rfim-frompapers,tkl,mypapers-frompapers,grants2010-frompapers,grants2013-hand,superstripes2014}

\end{document}